\documentclass[final, nopreprintline,12pt,3p,sort&compress]{elsarticle}
\usepackage{amssymb}
\usepackage[mathscr]{euscript}
\usepackage{amsthm}
\usepackage{multirow}

\usepackage{amsmath}
\usepackage{array}
\usepackage[hyphens]{url}
\usepackage{breakurl}
\usepackage{caption}

\usepackage[colorlinks,linkcolor=green, citecolor=green]{hyperref}

\journal{Arxiv}

\begin{document}

\begin{frontmatter}



\title{\textbf{Accelerated partial separable model using dimension-reduced optimization technique for ultra-fast cardiac MRI}}


\author[inst1]{Zhongsen Li\fnref{fn1}}

\affiliation[inst1]{organization={Center for Biomedical Imaging Research},
            addressline={Medical School, Tsinghua University}, 
            city={Beijing},
            postcode={100084}, 
            country={China}}
\affiliation[inst2]{organization={Institute of Science and Technology for Brain-Inspired Intelligence},
            addressline={Fudan University}, 
            city={Shanghai},
            postcode={200433}, 
            country={China}}
\author[inst2]{Aiqi Sun}
\author[inst1]{Chuyu Liu}
\author[inst1]{Haining Wei}
\author[inst1]{\\Shuai Wang}
\author[inst1]{Mingzhu Fu}
\author[inst1]{Rui Li\corref{cor1}}
\ead{leerui@tsinghua.edu.cn}
\cortext[cor1]{Corresponding author}


            


\begin{abstract}
\textit{Objective.} Imaging dynamic object with high temporal resolution is challenging in magnetic resonance imaging (MRI). Partial separable (PS) model was proposed to improve the imaging quality by reducing the degrees of freedom of the inverse problem. However, PS model still suffers from long acquisition time and even longer reconstruction time. The main objective of this study is to accelerate the PS model, shorten the time required for acquisition and reconstruction, and maintain good image quality simultaneously. \textit{Approach.} We proposed to fully exploit the dimension-reduction property of the PS model, which means implementing the optimization algorithm in subspace. We optimized the data consistency term, and used a Tikhonov regularization term based on the Frobenius norm of temporal difference. The proposed dimension-reduced optimization technique was validated in free-running cardiac MRI. We have performed both retrospective experiments on public dataset and prospective experiments on in-vivo data. The proposed method was compared with four competing algorithms based on PS model, and two non-PS model methods. \textit{Main results.} The proposed method has robust performance against shortened acquisition time or suboptimal hyper-parameter settings, and achieves superior image quality over all other competing algorithms. The proposed method is 20-fold faster than the widely accepted PS+Sparse method, enabling image reconstruction to be finished in just a few seconds. \textit{Significance.} Accelerated PS model has the potential to save much time for clinical dynamic MRI examination, and is promising for real-time MRI applications.
\end{abstract}



\begin{keyword}
Partial separable model \sep dynamic magnetic resonance imaging \sep dimension reduction \sep optimization algorithm \sep cardiac imaging
\end{keyword}

\end{frontmatter}


\section{Introduction}
\label{sec:intro}
Partial separable model (PS model)\cite{PSF_original_paper____LiangZP} is an important model for dynamic magnetic resonance imaging (MRI). PS model exploits the strong spatial-temporal correlations, and decomposes the dynamic images into several basis functions. Compared with other dynamic imaging models, such as k-t Sparse\cite{k-t_sparse_SENSE_FengLi} or L+S\cite{L+S}, the most prominent feature of the PS model is the explicit data dimension reduction. The reduction in the degrees of freedom lowers the sampling requirements, which enables higher temporal resolution to be achieved.

PS model can improve the temporal resolution of existing imaging applications, such as cardiac imaging\cite{PSF+LSQR____noECG_noBreathhold____realtime_rat_heart_imaging}\cite{rosenzweig2021_ungatedCMR_SSA_FARY}, delayed contrast enhancement (DCE) imaging\cite{PSF_xfL1_3D_var_density_stack_of_spiral_traj_sensitivity_encoding____U_V_and_sensMap_joint_optimization____liver_DCE}\cite{feng2023_4d_subsecond}, functional MRI\cite{PSF_LSQR_NUFFT_CG____multi_echo_spiral_in_Nav_spiral_out_Img____fMRI}\cite{mason2021subspace_fMRI}, speech imaging\cite{PSF+2spatial_TV+1temporal_TV____half_quadratic____Cartesian_Img+3Dcone_Nav____3D_vocal_imaging____recon_time_12h}\cite{PSF_xfL1____Spiral_Nav_Cartesian_Img____half_quadratic____big_FOV_full_vocal_imaging_recon_time_34min_each_slice}, and MR elastography\cite{mcilvain2022oscillate}. Moreover, PS model can bring new imaging techniques. For example, realtime phase contrast imaging\cite{PSF+xfL1_recon_ref____PSF+Usparse_recon_complex_difference____realtime_2D_and_3D_PC_flow_imaging____half_quadratic____recon_time_2D_16min_3D_1h}\cite{PSF_LSQR____2D-PC_single_direction_velocity____recon_time_10min}\cite{PSF_LSQR____realtime_4Dflow} uses PS model to resolve the beat-by-beat velocity variations of blood flow\cite{sun2022motion}. Besides, PS model has close connections with subspace-constrained reconstruction method, which is used by many advanced techniques, such as MR fingerprinting\cite{subspace_MRF}\cite{hu2021_MRFrecon_Hankel_subspace}\cite{cao2022optimizedMRF} and MR multitasking\cite{MR-multitasking_original}\cite{MR-multitasking_cardiac}\cite{cao2022free}. Therefore, study on the PS model will be valuable for many research fields.

Many aspects of the PS model have been studied. Different k-space trajectories\cite{PSF+radial____k-t_space_pseudo-inverse____rat_pulmonary_DCE}\cite{PSF_3D_stack_of_spiral____spiral_navigator____LSQR_in_k-space_and_gridding} were used to improve the PS model sampling efficiency. Eckart-Young approximation was used for evaluating the PS model expression ability\cite{PSF_theory____Eckart_Young_approximation____enough_sampling_can_recover_low_rank_matrix}. Singular value variations were depicted to help select the best model order\cite{PSF_heart_digital_phantom_simulation____select_the_best_model_order____according_to_the_max_difference_of_singular_value}. Besides, the PS model can be used in combination with other methods to improve the image quality, such as multi-channel parallel imaging methods\cite{PSF+sparse+parallel-joint-model____spatial_variable_rank____first_try_on_3D-PSimaging____realtime_cardiac_imaging}, sparsity priors\cite{PSF+xfL1____ADMM____cardiac_imaging____recon_time_11min}\cite{PSF_Atlas_sparse____half_quadratic____3D_vocal_imaging____recon_time_47h_4p5h}\cite{PSF_SIDWT_sparse____ADMM____realtime_cardiac}, structured low-rank models\cite{T2-shuffling}\cite{PSF____tSVD_dynamic_low_rank_tensor_constraints____ADMM____cardiac_perfusion}, and many other complicated models\cite{PSF+GRASP____use_radial_center_data_recon_low_res_image_to_get_V____temporalTV____non-linearCG____liver_DCE____recon_time_30min}\cite{PSF_model____U_decompose_to_dense_U1_and_sparse_U2____Gaussian_and_Laplacian____ADMM____realtime_cardiac}\cite{PSF_radial____manifold_learning_nonlinear_V____realtime_cardiac}\cite{PSF+kernelPCA_V____dynamic_cardiac_and_perfusion}. Most recently, deep neural network is integrated into PS model formulation to solve this problem in a data-driven manner\cite{PSF____unroll_network_U_and_V_joint_optimization}\cite{chen2022_nonCart_subspace_learn}\cite{cao2022_psNet}.

However, the PS model suffers from slow imaging speed. Because the dynamic process needs a relatively long period to reveal its spatial-temporal correlations\cite{PSF+kernelPCA_V____dynamic_cardiac_and_perfusion}, imaging methods based on the PS model usually needs long acquisition time and a large number of frames for reconstruction\cite{PSF_xfL1____Spiral_Nav_Cartesian_Img____half_quadratic____big_FOV_full_vocal_imaging_recon_time_34min_each_slice}\cite{PSF+xfL1_recon_ref____PSF+Usparse_recon_complex_difference____realtime_2D_and_3D_PC_flow_imaging____half_quadratic____recon_time_2D_16min_3D_1h}\cite{PSF_Atlas_sparse____half_quadratic____3D_vocal_imaging____recon_time_47h_4p5h}, which severely limits the clinical application of the PS model. Unfortunately, the acceleration of the PS model is still regarded as a limitation or future work in many papers\cite{PSF_LSQR____2D-PC_single_direction_velocity____recon_time_10min}\cite{PSF+GRASP____use_radial_center_data_recon_low_res_image_to_get_V____temporalTV____non-linearCG____liver_DCE____recon_time_30min}. To the best of our knowledge, there has been no literature which is dedicated to the acceleration problem of the PS model.

The main purpose of this work is to give a detailed analysis of the PS model optimization problem, and provide practical acceleration schemes to reduce the acquisition and reconstruction time. The key idea is the dimension-reduced optimization technique. We noticed that the dimension-reduction property of the PS model is not fully exploited by current reconstruction algorithms. We hypothesize that if the optimization is constrained in the low dimensional subspace, not only will the computation time be significantly reduced, but also the reconstruction performance will be more stable. Therefore, we optimized the data consistency term and carefully designed the regularization term for PS model. The proposed method is validated in free-running cardiac MRI. The results show that the total imaging time (including acquisition and reconstruction) can be reduced to only a few seconds by the proposed method, while good image quality is maintained.

This paper is organized as follows. Section II reviews the research background. Section III elaborates on the dimension-reduced optimization technique, with its theoretical proof and implementation details. Section IV provides the experimental settings. The results are shown in Section V and discussed in Section VI. Section VII concludes this paper.

\section{Research Background}
\label{sec:Background}

The signal equation of multi-channel dynamic MRI acquisition is written as:
\begin{equation}
y_{j}(\textbf{k},t)=\int  s_{j}(\textbf{r})\cdot x(\textbf{r},t)\cdot e^{-i2\pi \textbf{k}\cdot \textbf{r}}\cdot d\textbf{r}+e_j(\textbf{k},t),
\label{MRI_signal}
\end{equation}
where $x(\textbf{r},t)$ is the dynamic image function, subscript $j$ denotes the $j$-th coil, $s_j(\textbf{r})$ is the coil sensitivity, $y_{j}(\textbf{k},t)$ is the measured k-space data, and $e_j(\textbf{k},t)$ is the noise.

PS model exploits the partial separability of the dynamic data. 
\begin{equation}
x(\textbf{r},t)=\sum_{l=1}^{L} u_{l}(\textbf{r})\cdot v_{l}(t),
\label{xt_PS_model}
\end{equation}
where $u_{l}(\textbf{r})$ is the spatial basis functions and $v_{l}(t)$ is the temporal basis functions. If $N$ spatial locations and $T$ time frames are considered, $x(\textbf{r},t)$ can be formulated into a $N\times T$ Casorati matrix. The PS model can be written into a simple low-rank decomposition formula:
\begin{equation}
X=U\cdot V \,\,\,\, (U\in C^{N\times L}, V\in C^{L\times T}).
\label{X_UV}
\end{equation}
The MRI equation\eqref{MRI_signal} can be written as:
\begin{equation}
y=M\odot F\cdot S\odot (U\cdot V)+E,
\label{y}
\end{equation}
where $S$ is multi-channel coil sensitivity map, $F$ is 2D Discrete Fourier transform (DFT), $M$ is the $(\textbf{k},t)$ undersampling mask, $y$ is the multi-channel k-space data, and $E$ is the noise. ``$\cdot$'' indicates matrix multiplication, ``$\odot$'' indicates element-wise multiplication.

\begin{figure*}
\centerline{\includegraphics[width=\columnwidth]{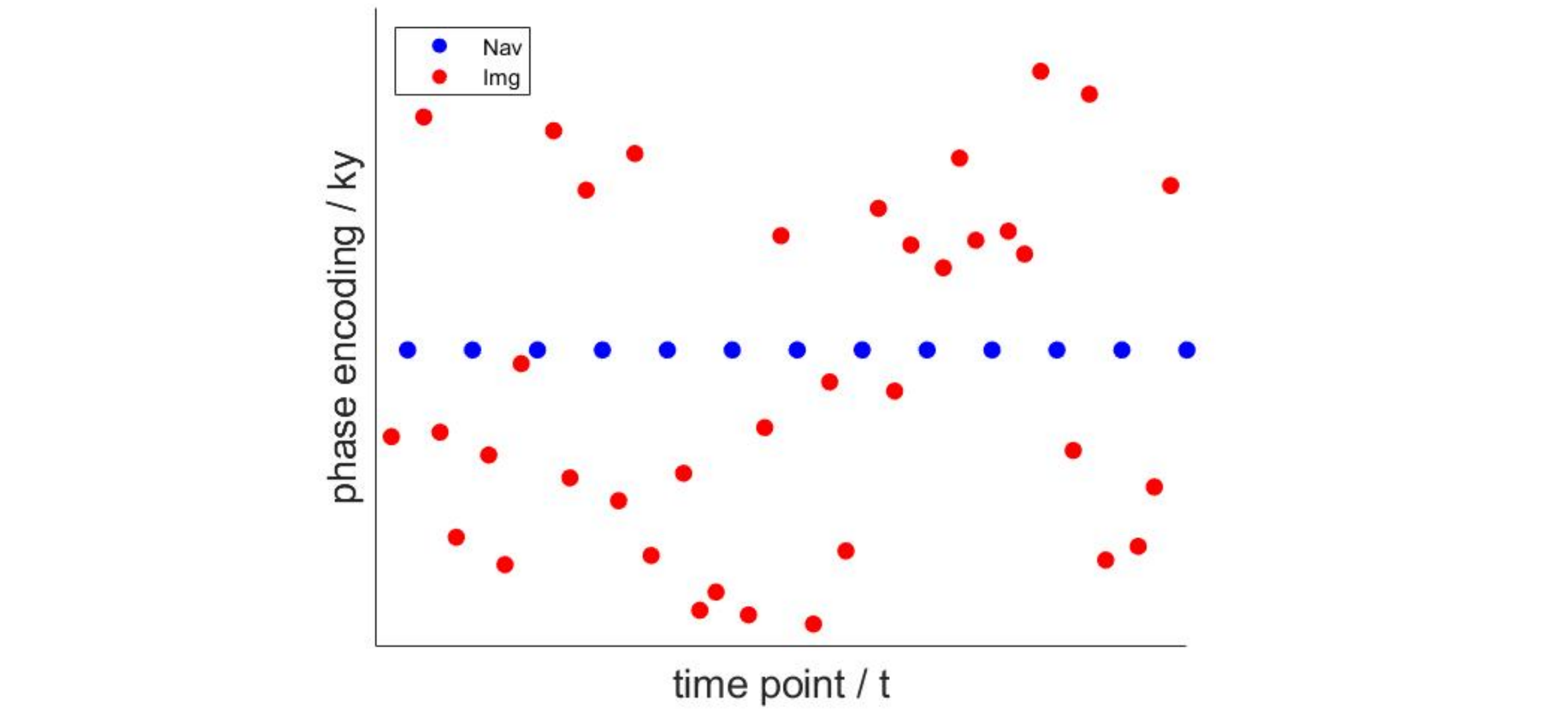}}
\vspace{-0.5em}
\caption{Interleaved (k-t) space sampling pattern for PS model. Because Cartesian trajectory is used, frequency encoding direction is fully-sampled, which is perpendicular to the paper and not depicted in the figure. The undersampling happens in the (ky-t) space, where ky is the phase encoding direction, t is the time dimension. The blue dots represent the ``Navigating data". The red dots represent the ``Imaging data".
}
\vspace{-1em}
\label{PSRTsampling}
\end{figure*}

The data acquisition of PS model is divided into two parts, as illustrated in ~\autoref{PSRTsampling}. ``Navigating data'' $y_{Nav}$ samples the center region of k-space at high temporal resolution, while ``Imaging data" $y_{Img}$ slowly samples the whole k-space. The two parts of data are sampled in an interleaved fashion. A two-step framework is adopted for image reconstruction. First, singular value decomposition (SVD) is performed on $y_{Nav}$ and the first $L$ singular vectors are extracted as $V$. Second, $V$ is assumed to be fixed, $U$ is calculated by minimizing the noise energy:

\begin{equation}
\begin{aligned}
V&=V_{Nav}(:,1:L)^{T}\,\,\,\,(y_{Nav}=U_{Nav}\cdot \Sigma \cdot V_{Nav}) \\
U&=\mathop{\arg\min}\limits_{U}\frac{1}{2} \left \| M\odot F\cdot S\odot (U\cdot V)-y_{Img} \right \| _2^2.
\end{aligned}
\label{lsqr_U}
\end{equation}

For simplicity, an encoding operator $A$ is defined for the PS model problem:
\begin{equation}
A(U)=AU=MFSVU.
\label{A}
\end{equation}
where $V$, $S$, $F$ and $M$ can be expressed as the left-multiplication form because they are all linear transformations to $U$. Then equation\eqref{lsqr_U} can be written into a least-square problem:
\begin{equation}
U=\mathop{\arg\min}\limits_{U}\frac{1}{2} \left \| AU-y_{Img} \right \| _2^2.
\label{PS-LSQR}
\end{equation}
In this paper, this model is named ``PS-LSQR". Besides, additional constraints can be incorporated into the PS model:
\begin{equation}
U=\mathop{\arg\min}\limits_{U}\frac{1}{2} \left \| AU-y_{Img} \right \| _2^2+\lambda R(WU),
\label{PS-R}
\end{equation}
where $W$ is a linear transform operator, $\lambda$ is the weighting coefficient, and $R$ is the penalty function. Equation\eqref{PS-R} is categorized as the ``PS-R" model.

\section{Proposed Method}
\label{sec:method}

\subsection{Dimension-Reduced Optimization}

The optimization variable of PS-R model in \eqref{PS-R} is the spatial basis $U \in C^{N\times L}$. Each row of $U$ is actually the projection coefficients onto the $L$-dimensional subspace $V$. According to the basic assumption of PS model, dynamic data has strong spatio-temporal correlations, which leads to $T\gg L$. However, this dimension-reduction property is not always maintained in the optimization process in previous works, which means much computation is wasted in high-dimensional space although we only seek a solution in the $L$-dimensional subspace. Therefore, the essence of dimension-reduced optimization technique is to constrain the computation process in the subspace.

The major computation of PS-R model is afforded by two parts. The first part is the $A^HA$ operator, which is derived from the first-order derivative of data consistency term. The second part is associated with the regularization term, which can be recognized by $W$ and $W^H$ operator in the iteration. These two parts need to be constrained in the subspace to derive a dimension-reduced optimization algorithm. 

\subsection{Dimension-Reduced $A^HA$ Operator}

Recalling the definition of $A$ operator in equation\eqref{A}, the $A^HA$ operator is defined by:
\begin{equation}
A^HA(U)=A^HAU=V^HS^HF^HM^HMFSVU,
\label{AHA_2M}
\end{equation}
where $M$ is a diagonal matrix and only contains 0-1 real values, thus $M^H=M$. Therefore:
\begin{equation}
A^HA(U)=A^HAU=V^HS^HF^HMFSVU.
\label{AHA_orig}
\end{equation}
We need two theorems to optimize the $A^HA$ operator:

\subsubsection{\rm \textbf{Theorem 1: operator exchange-ability}}
The computation order of operator $V$ can be exchanged with $S$ and $F$. (The proof is provided in Appendix A). With Theorem 1, we can rewrite the $A^HA$ operator as:
\begin{equation}
A^HA(U)=A^HAU=S^HF^HV^HMVFSU.
\label{AHA_VHMV}
\end{equation}
Note that before exchanging the operators, the coil sensitivity weighting $S$ and 2D DFT transform will be performed for a total of $T$ frames of images. However, after exchanging the operators, it needs only to perform the same operation directly on $U$, for a total of $L$ frames of images. The computational complexity of the $S$ and $F$ operators is reduced by $\frac{T}{L}$.

\subsubsection{\rm \textbf{Theorem 2: operator merge-ability}}
The three operators $V^H$, $M$ and $V$ can be merged into a single operator $\Phi \in C^{N\times L\times L}$. (The proof is provided in Appendix B). With Theorem 2, equation\eqref{AHA_VHMV} can be simplified to:
\begin{equation}
A^HA(U)=A^HAU=S^HF^H\Phi FSU,
\label{AHA_Phi}
\end{equation}
where $\Phi \in C^{N\times L\times L}$ is composed of $N$ unique matrices of shape $L\times L$ for different spatial locations. The computational complexity of $A^HA$ operator is reduced by $\frac{T}{L}$. This optimization is irrelevant of $R$ functions, thus can benefit all kinds of subspace-constrained algorithms.

\subsection{Dimension-Reduced Regularization Term}

The dimension-reduced property of the regularization term is highly dependent on the $R$ function. For example, if L1-norm in the temporal Fourier domain is used for regularization, the data must be transformed to the $(\textbf{r},f)$ domain (spatial-frequency domain) to enforce sparsity, which will elevate the data dimension. In order to maintain the dimension-reduced property, we propose the following model based on the generalized Tikhonov regularization:
\begin{equation}
U=\mathop{\arg\min}\limits_{U}\frac{1}{2} \left \| AU-y_{Img} \right \| _2^2+\frac{\lambda}{2}  \left \|  DVU\right \| _F^2,
\label{PS-TDF}
\end{equation}
where $D$ is the finite difference operator along the time dimension.

The proposed PS-R model is designed based on the following observations. First, the proposed method uses the Frobenius-norm of the temporal gradient as the regularization, which presumes that the signal changes smoothly with time. This regularization shares the same prior assumption as the sparsity in $(\textbf{r},f)$ space, so we hypothesize that the proposed method can achieve comparable image quality with the PS+Sparse model\cite{PSF+xfL1____ADMM____cardiac_imaging____recon_time_11min}. Second, the proposed model can be solved by the following low-dimensional linear equations:
\begin{equation}
\begin{aligned}
U&=(A^HA+\lambda \Psi )^{-1}A^Hy_{Img} \\
\Psi &=V^HD^HDV \in C^{L\times L}.
\label{solve_PS-TDF}
\end{aligned}
\end{equation}
Because the $A^HA$ operator has already been optimized to be constrained in subspace, and $\Psi \in C^{L\times L}$ is also low-dimensional, this PS-R model can be solved very efficiently.

\subsection{A Summary of Comparative Algorithms}

We implemented five PS model algorithms for comparison:
\begin{itemize}
    \item {PS-LSQR}
    \begin{equation}
    \begin{aligned}
    U=\mathop{\arg\min}\limits_{U}\frac{1}{2} \left \| AU-y_{Img} \right \| _2^2.
    \end{aligned}
    \label{sum_PS-LSQR}
    \end{equation}
    This is an unconstrained least-square model without any regularization term. This model is the simplest PS model, but will become severely ill-posed if the acquisition time is shortened. Therefore, PS-LSQR is implemented as the baseline algorithm.
    
    \item {PS-xfL1}
    \begin{equation}
    \begin{aligned}
    U=\mathop{\arg\min}\limits_{U}\frac{1}{2} \left \| AU-y_{Img} \right \| _2^2+\lambda \left \|  F_t(VU)\right \| _1,
    \end{aligned}
    \label{sum_PS-xfL1}
    \end{equation}
    where $F_t$ is Fourier transform along the time dimension. This is a widely accepted PS+Sparse model \cite{PSF+xfL1____ADMM____cardiac_imaging____recon_time_11min}, which exploits the sparsity of the Fourier coefficients of dynamic images. However, this model will destroy the dimension-reduced property in the optimization process because the sparsity must be enforced in the $(\textbf{r},f)$ domain.
    
    \item {PS-LLR}
    \begin{equation}
    \begin{aligned}
    U=\mathop{\arg\min}\limits_{U}\frac{1}{2} \left \| AU-y_{Img} \right \| _2^2+\lambda \sum_{i}^{} \left \| (T_{LLR}U)_i \right \| _*,
    \end{aligned}
    \label{sum_PS-LLR}
    \end{equation}
    where $T_{LLR}$ is a linear transform which extracts small patches from the image, and $\left \|  \right \| _*$ is the nuclear norm. This is a  PS+Locally-Low-Rank model\cite{T2-shuffling}, which exploits the local similarity of the spatial basis images. For each step, this model needs to compute SVD to calculate the singular values. Besides, the $T_{LLR}$ transform requires a random shift offset for each step, which causes that the convergence of this model is intrinsically slower than those non-random algorithms.

    \item {PS-SIDWT}
    \begin{equation}
    \begin{aligned}
    U=\mathop{\arg\min}\limits_{U}\frac{1}{2} \left \| AU-y_{Img} \right \| _2^2+\lambda \left \|  W_sF_t(VU)\right \| _1,
    \end{aligned}
    \label{sum_PS-SIDWT}
    \end{equation}
    where $F_t$ is Fourier transform along the time dimension, and $W_s$ is shift-invariant discrete wavelet transform (SIDWT)\cite{ref_MODWT}. This is a recently proposed PS+Sparse model which utilizes the strong sparsity of the SIDWT coefficients\cite{PSF_SIDWT_sparse____ADMM____realtime_cardiac}. However, this model also destroys the dimension-reduced property because the sparsity is enforced in a high-dimensional domain. Besides, SIDWT is more time-consuming than DFT.
    
    \item {proposed}
    \begin{equation}
    U=\mathop{\arg\min}\limits_{U}\frac{1}{2} \left \| AU-y_{Img} \right \| _2^2+\frac{\lambda}{2}  \left \|  DVU\right \| _F^2,
    \label{sum_PS-TDF}
    \end{equation}
    where $D$ is the finite difference operator. The regularization term is designed to be simple to avoid time-consuming computations, but also effective in improving the inverse problem conditioning. This model is a generalized Tikhonov regularization problem, which can be solved very efficiently by low-dimensional linear equations.
    
\end{itemize}

For PS-LSQR and the proposed model, conjugate gradient (CG) solver is used as the optimizer. For PS-xfL1, PS-LLR, and PS-SIDWT model, Project Onto Convex Set (POCS) algorithm is used as the optimizer. The $\lambda $ is the only one hyper-parameter in POCS algorithm, which avoids the complicated impact of multiple hyper-parameters, so that a fair comparison can be performed on different models.

Besides, two non-PS model methods for dynamic MRI reconstruction are also implemented in this work. The two methods directly solve for the dynamic image $X$. As defined above, $F$ is 2D DFT, $S$ is coil sensitivity map and $M$ is (k-t) undersampling mask.
\begin{itemize}
    \item {L+S}
    \begin{equation}
    \begin{aligned}
    X,L,S=\mathop{\arg\min}\limits_{X,L,S}\frac{1}{2} \left \| MSFX-y_{Img} \right \| _2^2+\lambda_1 \left \| F_tS \right \| _{1}+\lambda_2 \left \| L \right \| _*+\lambda_3 \left \| L+S-X \right \| _2^2.
    \end{aligned}
    \label{sum_L+S}
    \end{equation}
    This model\cite{L+S_2022IEEEACCESS} is a variant of the $L+S$ model\cite{L+S}, which decomposes the data into a low-rank component and a sparse component. In this model, Fourier transform along the time dimension $F_t$ is applied to the $S$ component to enhance its sparsity, and quadratic term is used to relax the constraints. $L+S$ is a popular reconstruction model recently due to its flexibility and interpretability \cite{li2021_LplusS_T1mapp}\cite{wang2020_LplusS_TGV_lowRank}\cite{sun2022_LplusS_slice_lowrank}\cite{daniel2023_LplusS_BlochSimu_T2map}. However, $L+S$ model doubles the size of variables, which will be computational-expensive for large-scale problems.

    \item {TNN}
    \begin{equation}
    \begin{aligned}
    X,H=\mathop{\arg\min}\limits_{X,H}\frac{1}{2} \left \| MSFX-y_{Img} \right \| _2^2+\lambda_1 \left \| H \right \| _{*}+\lambda_2 \left \| R_HFX-H \right \| _2^2.
    \end{aligned}
    \label{sum_TNN}
    \end{equation}
    This is a tensor low-rank model\cite{TNN_paper}. $R_H$ is a sliding window operator, which extract patches from 3D data to construct a Hankel matrix\cite{liu2021calibrationless}\cite{zhao2022calibrationless}. $H$ is an auxiliary variable, which is used for approximating the Hankel matrix in k-space. Here, ``$\left \| \cdot \right \|_*$'' indicates the tensor nuclear norm (TNN)\cite{kolda2009tensor}, which is calculated by high order SVD (HOSVD)\cite{de2000multilinear}, $H=C*K_1*K2*K3$. In this formula, $C$ is the core tensor, $K_1$, $K_2$, and $K_3$ are factor matrices along each tensor-mode, ``$*$'' is the tensor-tensor product operator. Tensor low-rank model can exploit the multi-dimensional correlations of the dynamic data\cite{zhang2022_TNN_and_CasoratiNN}\cite{cui2021_weighted_TNN}, at the cost of high computational complexity. 
\end{itemize}

\begin{figure*}
\centerline{\includegraphics[width=\columnwidth]{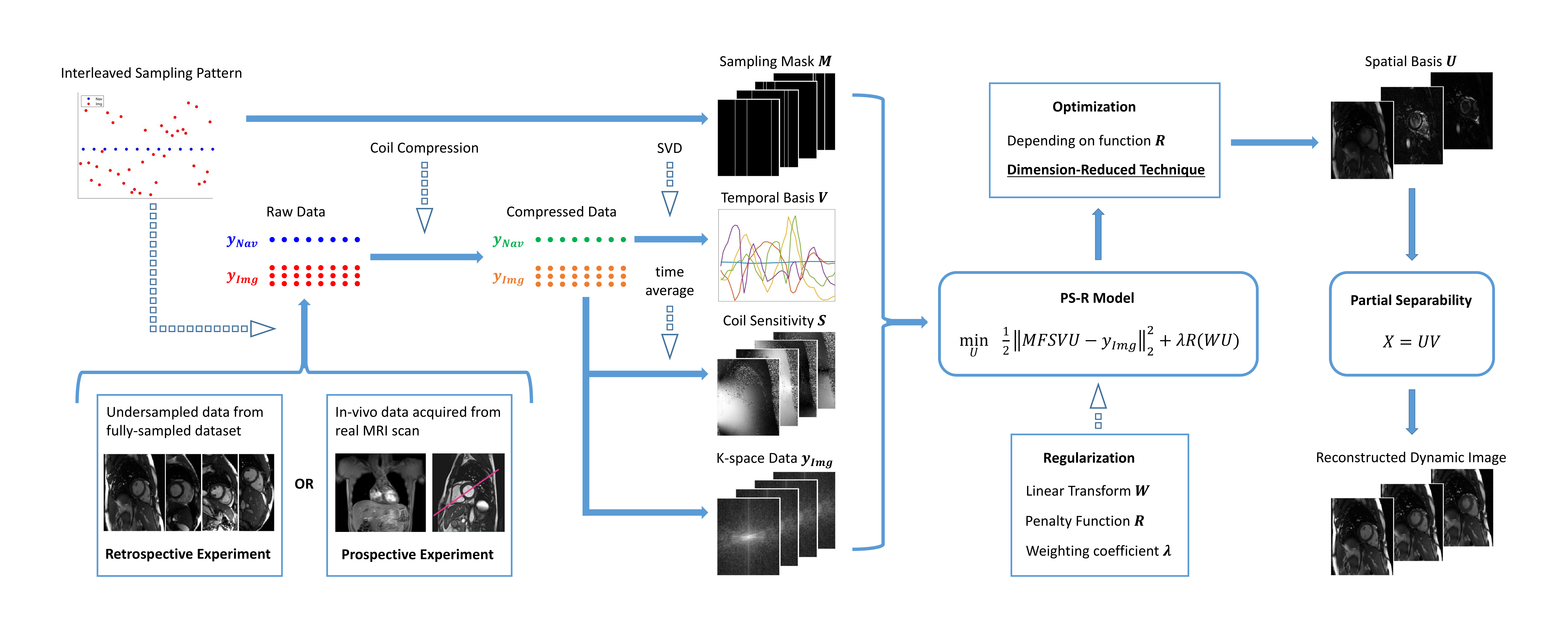}}
\vspace{-0.5em}
\caption{The image reconstruction pipeline for PS model algorithms. The simulation data and in-vivo data are both reconstructed by this pipeline. Firstly, coil compression is applied to the input data to reduce computation complexity. Then, (k-t) sampling mask $M$, temporal basis functions $V$, coil sensitivity maps $S$ and undersampled k-space data $y$ are calculated from the compressed data. The regularization term and optimization algorithm are designed independently for different PS model algorithms. When the solution $U$ is obtained, the reconstructed images are calculated by $X=U\cdot V$.}
\vspace{-1em}
\label{pipeline}
\end{figure*}

\section{Experiments}
\label{sec:experiments}

\subsection{Selection of the Model Order}

The model order $L$ needs to be determined first for PS model algorithms. In this work, we adopted a simulation method for determining the best model order $L$\cite{T2-shuffling}. In this method, the observation noise is assumed to be independent complex Gaussian random variables:
\begin{equation}
e(\textbf{r},t)\sim N(0,\sigma ^2),
\label{noise_model}
\end{equation}
where $\sigma^2$ is the noise variance. For the PS model, the expectation for the pixel-wise relative reconstruction error $E_{pix}$ can be estimated by the following equation\cite{T2-shuffling}:
\begin{equation}
E_{pix}=\sqrt{\frac{1}{N}\cdot E\left \{ \left \| \widehat{X}-X  \right \|_F^2  \right \}} =\sqrt{\left \| X\cdot V^H\cdot V-X  \right \|_F^2+L\cdot\sigma^2},
\label{noise_model}
\end{equation}
where $\widehat{X}$ is the reconstructed image, $X$ is the ground-truth image, $E\left \{ \cdot  \right \}$ means taking the expectation of random variable.

Using this equation, a simulation study is performed on the OCMR dataset\cite{OCMR}, which is a public dataset containing multi-coil k-space data of cardiac MRI. The noise level $\sigma^2$ is approximated by the intensity variance $\sigma_0^2$ calculated from the empty region in the images. The reconstruction error $E_{pix}$ is calculated for different $L$ values ($L=1\sim 100$) for each data case. The mean $E_{pix}$ is plotted to $L$, which produces the $E_{pix}\sim L$ error curve. Besides, $\sigma^2$ is also changed to obtain the error curves under different noise levels. Seven noise levels ($\sigma^2=0.1\sigma_0^2, 0.2\sigma_0^2, 0.5\sigma_0^2, \sigma_0^2, 2\sigma_0^2, 5\sigma_0^2, 10\sigma_0^2$) are simulated in this study. The best $L$ selected from this study is used in the following experiments.

\begin{figure*}
\centerline{\includegraphics[width=\columnwidth]{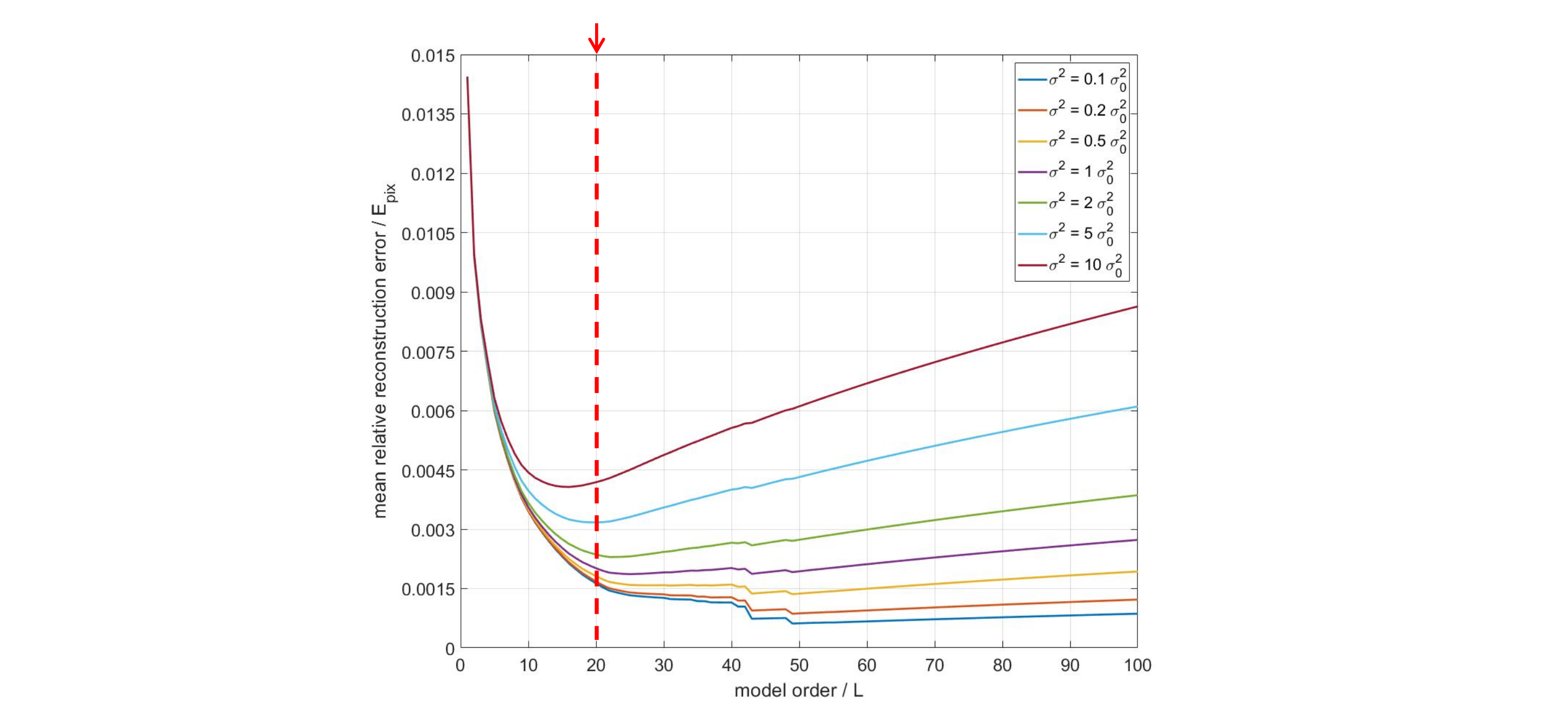}}
\vspace{-0.5em}
\caption{The $E_{pix}\sim L$ curves obtained from the simulation study. The horizontal axis is the model order $L$, the vertical axis is the mean relative reconstruction error $E_{pix}$. The reconstruction error is simulated under seven noise levels. The red arrow points at the model order $L=20$, which is used for all PS model algorithms in this work.}
\vspace{-1em}
\label{Lcurve_new}
\end{figure*}

\subsection{Reconstruction Pipeline}

An image reconstruction pipeline is established to provide a fair platform for algorithm comparison, as shown in ~\autoref{pipeline}. Simulation data and in-vivo data are both reconstructed by this pipeline. Firstly, input data is undersampled according to the interleaved pattern. Afterwards, the coil channel number is reduced to 6 by GCC coil compression method\cite{GCC_coil_compression}. Then, the data is normalized by its maximum magnitude. Then, SVD is performed on ``Navigating data'' to extract temporal basis functions $V$. ``Imaging data'' is averaged along the time dimension and used for estimating coil sensitivity maps $S$. Undersampling mask $M$ and k-space data $y$ are also calculated from the ``Imaging data''. The regularization term and optimization algorithm are designed independently for different PS models, as described in Section III. When the solution $U$ is obtained, the dynamic images are calculated by $X=U\cdot V$. All the computations were performed in MATLAB (The MathWorks, Natick, MA) in a personal computer equipped with 96 GB RAM and an Intel i9-10900F CPU.

\subsection{Retrospective Experiments}

Retrospective experiments were performed on OCMR dataset\cite{OCMR}. Only fully-sampled data was used to provide reference images. To control the variables for better comparison, we selected the data cases of the same spatial resolution and spatial coverage at short-axis orientation. Besides, edge slices with little or no motion were excluded from the dataset, because these slices are always reconstructed in high quality. A total of 49 slices from 17 subjects were eventually included in the retrospective experiments.

\begin{figure*}
\centerline{\includegraphics[width=\columnwidth]{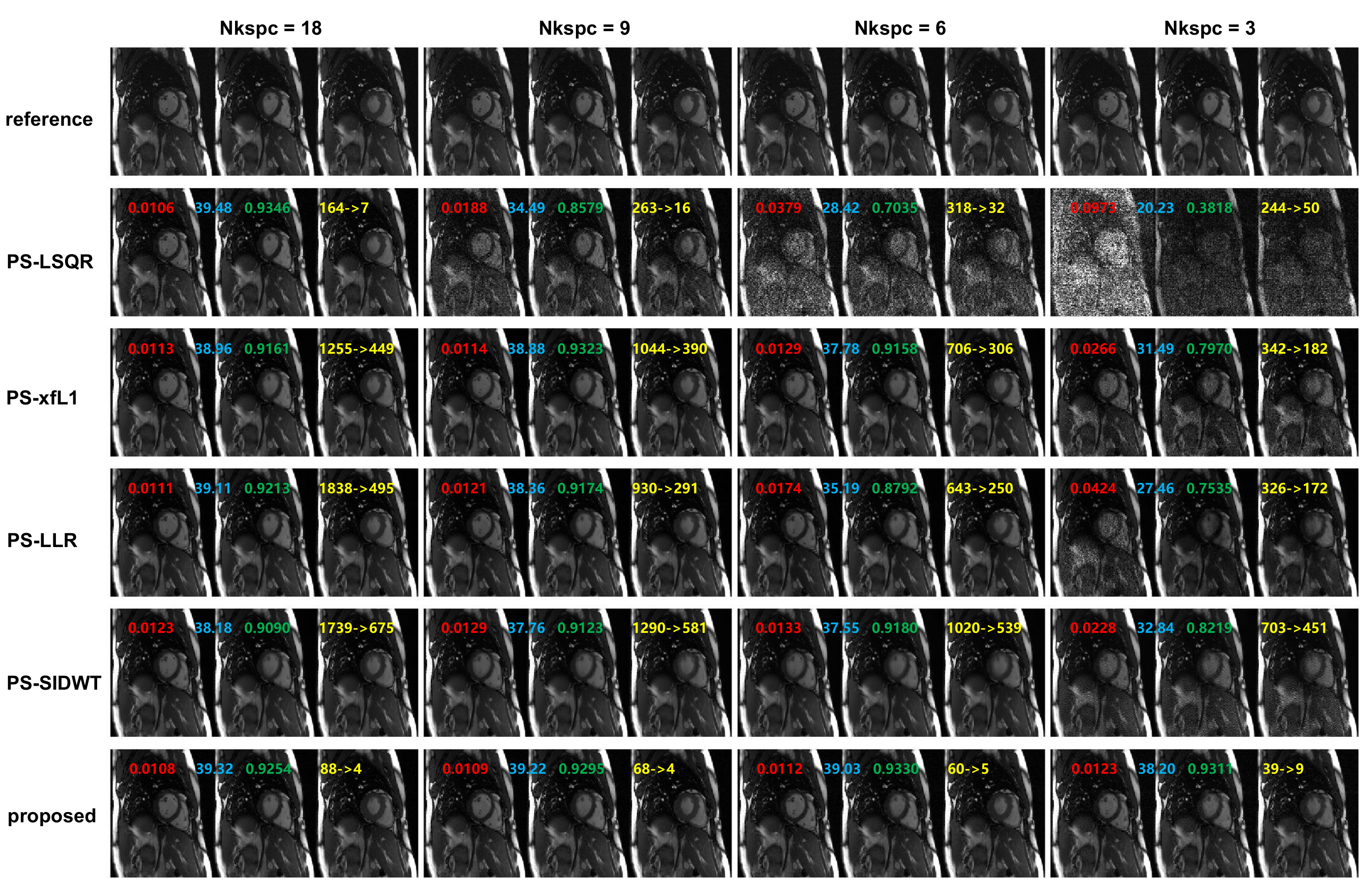}}
\vspace{-0.5em}
\caption{Reconstructed images for different Nkspc settings of one case data. The image frames at the time of diastole, cardiac contraction and systole are displayed. Quantitative metrics are calculated based on the reference image (red: nRMSE, blue: PSNR, green: SSIM). The reconstruction time is marked in yellow color. For each image, the reconstruction time before $A^HA$ operator optimization is marked on the left of the arrow, the reconstruction time after $A^HA$ operator optimization is marked on the right of the arrow.}
\vspace{-1em}
\label{retro-1__Nkspc}
\end{figure*}

Simulation data were generated by the following steps. First, the dynamic k-space data was interpolated to a given temporal resolution, which was set to be approximately 12 ms in this paper, in order to be consistent with the settings used for in-vivo data acquisition. Second, the raw data of one cardiac cycle was repeatedly extended to a given total length. In order to avoid complete periodicity of the simulated data, we performed random interpolation along the time dimension among cardiac cycles to simulate the variable heart beats. Third, each data was undersampled according to the interleaved sampling pattern as shown in ~\autoref{PSRTsampling}. The generated data was used for the following reconstruction experiments.

\subsubsection{Comparative Study on Nkspc}

The quality of PS model reconstruction is dependent on the total amount of  ``Imaging data''. More ``Imaging data'' leads to higher reconstruction quality, while at the expense of longer acquisition time. In this study, we use Nkspc to quantify the ``Imaging data'' amount, which denotes the number of full k-space if we neglect the phase encoding position and collect all the ``Imaging data'' compactly. Compared with total acquisition time, Nkspc is a better measurement of the difficulty for solving a PS model problem, because Nkspc is irrelevant of sequence TR, scan resolution, spatial coverage, and many other practical factors. Reconstruction experiments were performed on the retrospective dataset under Nkspc=18, 9, 6, 3. Quality metrics of nRMSE(normalized root mean square error), PSNR(peak signal to noise ratio), SSIM(structural similarity) were calculated based on the fully-sampled reference image, and the reconstruction time was recorded.

\subsubsection{Comparative Study on $\lambda $}

A brute-force search for the best $\lambda $ was performed for each PS model algorithm. The search range of regularization coefficient $\lambda $ was fixed as $\left [ 0.01:0.01:0.09,\,0.1:0.1:1 \right ]$ for Nkspc=9, $\left [ 0.05:0.01:0.09,\,0.1:0.1:0.9,\,1:1:5 \right ]$ for Nkspc=6, and $\left [ 0.1:0.1:0.9,\,1:1:10 \right ]$ for Nkspc=3. For PS-xfL1, the search range was multiplied by $1\times 10^{-5}$. For PS-LLR, the search range was multiplied by $1\times 10^{-4}$. For PS-SIDWT, the search range was multiplied by $1\times 10^{-5}$. For the proposed model, the search range was multiplied by $1\times 10^{-1}$. Scaling of the search range guarantees that $\lambda $ is suitable for each algorithm. Image reconstruction was performed for each combination of $\lambda $ and Nkspc values. Image quality metrics of nRMSE, PSNR, SSIM and the reconstruction time were recorded.

\subsection{Prospective Experiments}

A free-running cardiac imaging sequence based on the interleaved sampling pattern was implemented on a 3T MRI scanner (Philips IngeniaCX R5.7.1; Best, Netherlands). The sequence is based on balanced steady state free precession (bSSFP) sequence, which is the standard sequence for heart function assessment in clinical practice. No ECG device nor respiratory-monitoring equipment is needed for this sequence. The detailed sequence parameters are: TR=$3.4ms$, TE=$1.7ms$,  flip angle=$45^{\circ}$, FOV=$300mm\times 300mm$, voxel size=$2mm\times 2mm$, matrix size=$150\times 150$, one ``Navigating data'' readout is acquired for every three ``Imaging data'' readouts, the temporal resolution is approximately $13ms$.

Four healthy volunteers were recruited for validating the proposed method. The in-vivo experiments were approved by the Institution Review Board of the local institution. Written informed consent was obtained from the volunteers. The best $\lambda $ selected from the retrospective experiments was used for each algorithm. Other reconstruction parameters and procedures were held exactly the same as in retrospective experiments.

\begin{figure*}
\centerline{\includegraphics[width=\columnwidth]{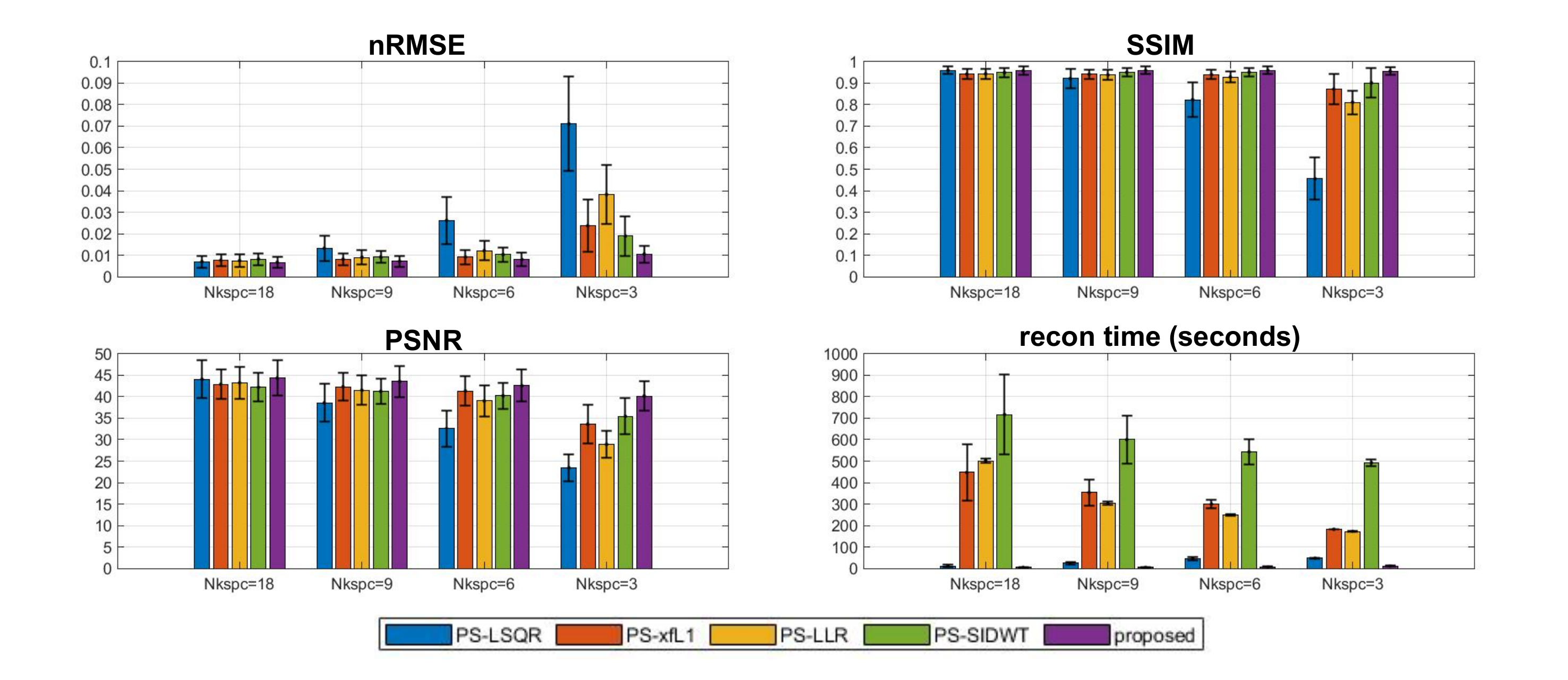}}
\vspace{-0.5em}
\caption{Quantitative metrics for different PS model reconstruction algorithms under different Nkspc settings. The metric values are calculated on all the data cases in the retrospective dataset. For each metric, the bar height indicates the mean value on the dataset, the error stick on each bar indicates the standard deviation (std) among different data cases.}
\vspace{-1em}
\label{retro-2__Nkspc__barPlot}
\end{figure*}

\subsection{Comparative Experiments with non-PS model Methods}
The proposed method is compared with two non-PS model methods, L+S\cite{L+S_2022IEEEACCESS} and TNN\cite{TNN_paper}. Both methods directly solve for the dynamic image, without explicit data dimension reduction. The experiment settings are held the same as the settings for the PS model methods. Parameter-tuning has been performed for these two methods, and the best result of each algorithm is used for comparison. Besides, the results of zero-filling (ZF) reconstruction and PS-LSQR are also presented in this experiment for comparison. ZF reconstruction is the simplest reconstruction method without any modeling, which just fills the undersampled k-sapce data points with zeros.

\section{Results}

\subsection{Selection of the Model Order}

The $E_{pix}\sim L$ curves obtained from the simulation study are shown in ~\autoref{Lcurve_new}. It can be observed that when the model order $L$ increases, the reconstruction error curve will decrease first, and then will increase when $L$ reaches certain turning point. Obviously, the turning point is exactly the best model order in theory.

However, the position of the turning point is different for different noise levels. For low noise levels, the turning point is larger, which means greater $L$ is more preferable for reconstruction. For high noise levels, the turning point is smaller, which means smaller $L$ is better for reconstruction. In practice, the noise variance $\sigma^2$ varies with the MRI hardware and environment conditions. As is shown ~\autoref{Lcurve_new}, a good balance between model performance and noise amplification is achieved at $L=20$ for a wide range of noise levels. Therefore, $L=20$ is selected as the model order in this work, which is used for all the PS model algorithms in this work.

\subsection{Retrospective Experiments}

\subsubsection{Comparative Study on Nkspc}

The reconstructed images for different Nkspc settings are shown in ~\autoref{retro-1__Nkspc}. Under Nkspc=18, all the models produce similarly high quality images, with nRMSE slightly above 0.01, PSNR above 38 and SSIM above 0.90. When Nkspc is reduced to 9 and 6, the image quality of PS-LSQR model deteriorates obviously, while the performance of the other algorithms are only slightly affected. When the Nkspc is further reduced to 3, a notable image quality decline can be observed for the PS-xfL1, PS-LLR and PS-SIDWT methods. Only the proposed method can still keep good visual image quality. This suggests that the proposed method is the most stable algorithm when the acquisition time is significantly shortened. Besides, it can be observed that the $A^HA$ operator optimization significantly reduces the computation time for all algorithms. The acceleration factor increases with the Nkspc value, which means more reconstruction time can be saved if the acquisition time is prolonged. For all Nkspc settings, the proposed method is always the fastest algorithm.

The statistical results on the retrospective dataset are shown in ~\autoref{retro-2__Nkspc__barPlot}. The proposed method usually achieves the lowest average nRMSE, the highest average PSNR and SSIM. Moreover, the proposed method also shows the lowest standard deviation (std) nearly among all metrics, which implies that the proposed method has high robustness among different data cases. Additionally, the advantage of the proposed method in reconstruction speed is very prominent. In short, the proposed method achieves the best image quality and the fastest speed on the dataset, and is robust against shortened acquisition time.

\begin{figure*}
\centerline{\includegraphics[width=\columnwidth]{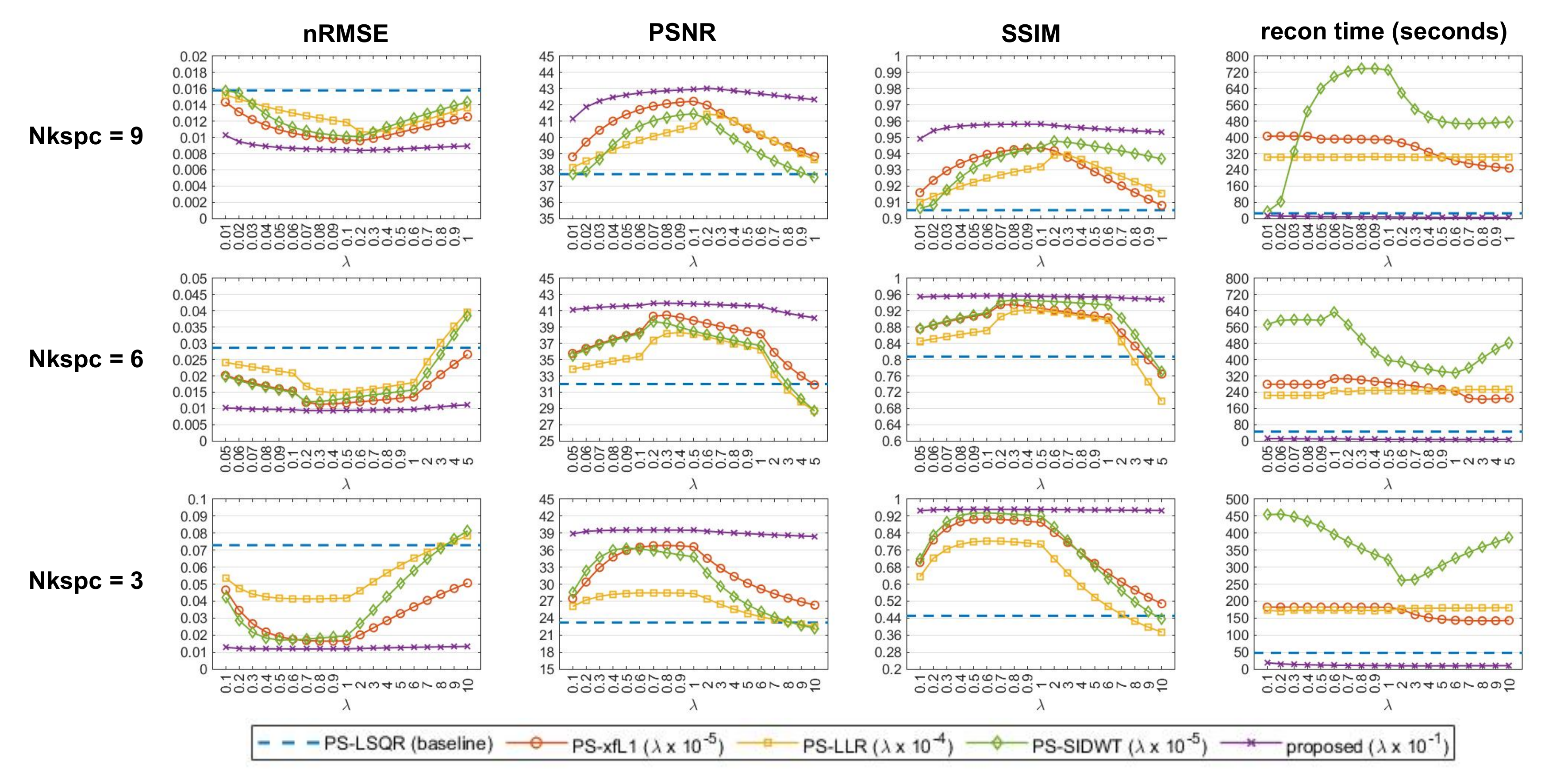}}
\vspace{-0.5em}
\caption{Results of the brute force search for the best $\lambda $ values for different PS models. Reconstruction is performed on all the cases in the dataset for each combination of $\lambda $ and Nkspc values. The mean value of the evaluation metrics is plotted to $\lambda $. The curves of PS-LSQR is plotted as a horizontal dotted line (blue color) for reference.}
\vspace{-1em}
\label{retro-3__Lambda__curve}
\end{figure*}

\subsubsection{Comparative Study on $\lambda $}

The results about the search for the best $\lambda $ is shown in ~\autoref{retro-3__Lambda__curve}. The best $\lambda $ has already been discovered for each model. The proposed method has the lowest mean nRMSE, the highest mean PSNR, the highest mean SSIM, and the lowest mean reconstruction time at each given $\lambda $. This indicates that the proposed method outperforms other models in a systematic level. Besides, when Nkspc is reduced, the gap between the curve peak of proposed method and other models increases, which means the proposed method displays stronger superiority when the acquisition time is shortened. Finally, the curve of proposed method is relatively flat compared with other methods, which implies the proposed method has stronger robustness against the hyper-parameter $\lambda $.

Comprehensively considering the quantitative metrics, the best $\lambda $ is chosen for each model. (For Nkspc=9, $\lambda = 1\times 10^{-6}$ for PS-xfL1, $\lambda = 2\times 10^{-5}$ for PS-LLR, $\lambda = 1\times 10^{-6}$ for PS-SIDWT, and $\lambda = 2\times 10^{-2}$ for the proposed method. For Nkspc=6, $\lambda = 3\times 10^{-6}$ for PS-xfL1, $\lambda = 4\times 10^{-5}$ for PS-LLR, $\lambda = 2\times 10^{-6}$ for PS-SIDWT, and $\lambda = 3\times 10^{-2}$ for the proposed method. For Nkspc=3, $\lambda = 7\times 10^{-6}$ for PS-xfL1, $\lambda = 7\times 10^{-5}$ for PS-LLR, $\lambda = 5\times 10^{-6}$ for PS-SIDWT, and $\lambda = 5\times 10^{-2}$ for the proposed method). Detailed quantitative metrics at the the best $\lambda $ value are listed in ~\autoref{table_best_lambda}.

\begin{figure*}
\centerline{\includegraphics[width=\columnwidth]{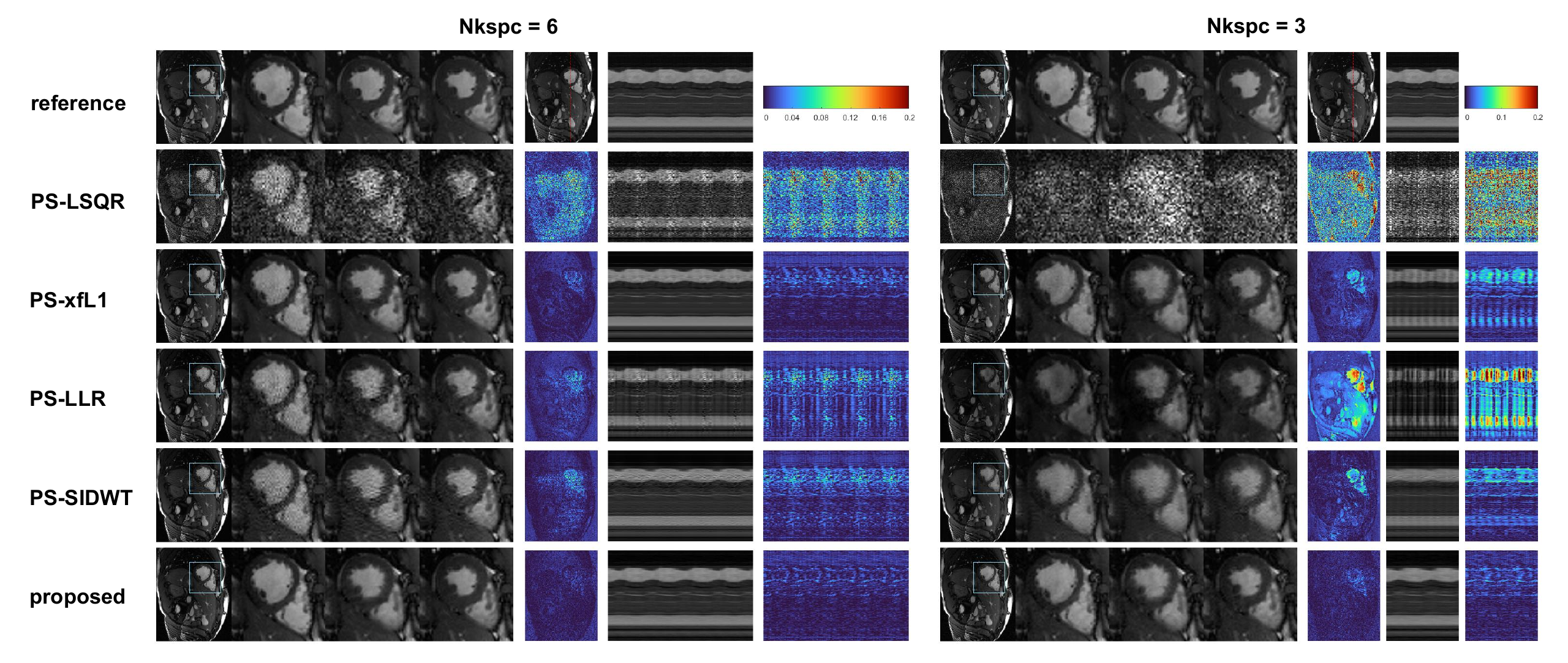}}
\vspace{-0.5em}
\caption{Reconstructed images by different PS models using the best $\lambda $. Reference images are shown in the first row. Reconstructed images of PS model algorithms are shown in other rows. The results under Nkspc=6 and Nkspc=3 are compared. For each Nkspc setting, the first column displays the reconstructed image. The second, third and fourth column display the zoomed image in the heart region at the time of diastole, cardiac contraction and systole, respectively. The fifth column displays the error map. Besides, a vertical red dotted line is also drawn across the heart right atrium. The 1D motion profile (which is called ``M-mode image'') at this line is plotted in the sixth column. The error map of the M-mode image is displayed at the seventh column. The error colorbar is plotted at the upper right corner.}
\vspace{-1em}
\label{retro-4__Lambda__ZOOM__Mmode}
\end{figure*}

\begin{table}[htbp]
\vspace{-0.5em}
  \centering
  \caption{Quantitative metrics of different PS models using the best $\lambda $}
  \resizebox{\linewidth}{!}{
    \begin{tabular}{cccccc}
    \hline
    Nkspc & Model & nRMSE & PSNR  & SSIM  & recon time (s) \\
    \hline
    \multirow{5}[2]{*}{9} 
          & PS-LSQR & 0.0132$\pm $0.0060 & 38.55$\pm $4.41 & 0.9214$\pm $0.0453 & 24.4$\pm $7.2 \\
          & PS-xfL1 & 0.0075$\pm $0.0029 & 43.20$\pm $3.77 & 0.9512$\pm $0.0198 & 388.8$\pm $20.4 \\
          & PS-LLR & 0.0083$\pm $0.0034 & 42.40$\pm $4.00 & 0.9474$\pm $0.0218 & 303.7$\pm $8.9 \\
          & PS-SIDWT & 0.0101$\pm $0.0027 & 41.45$\pm $3.77 & 0.9439$\pm $0.0216 & 732.4$\pm $29.4 \\
          & proposed & \textbf{0.0070}$\pm $\textbf{0.0026} & \textbf{43.83}$\pm $\textbf{3.78} & \textbf{0.9614}$\pm $\textbf{0.0170} & \textbf{5.7}$\pm $\textbf{0.9} \\
    \hline
    \multirow{5}[2]{*}{6} 
          & PS-LSQR & 0.0262$\pm $0.0110 & 32.55$\pm $4.29 & 0.8233$\pm $0.0799 & 45.7$\pm $6.9 \\
          & PS-xfL1 & 0.0092$\pm $0.0033 & 41.34$\pm $3.38 & 0.9398$\pm $0.0212 & 300.2$\pm $17.8 \\
          & PS-LLR & 0.0120$\pm $0.0049 & 39.23$\pm $3.86 & 0.9304$\pm $0.0261 & 249.2$\pm $5.1 \\
          & PS-SIDWT & 0.0121$\pm $0.0029 & 39.67$\pm $3.41 & 0.9432$\pm $0.0292 & 476.6$\pm $28.2 \\
          & proposed & \textbf{0.0080}$\pm $\textbf{0.0030} & \textbf{42.68}$\pm $\textbf{3.72} & \textbf{0.9607}$\pm $\textbf{0.0170} & \textbf{7.9}$\pm $\textbf{1.4} \\
    \hline
    \multirow{5}[2]{*}{3} 
          & PS-LSQR & 0.0711$\pm $0.0218 & 23.43$\pm $3.06 & 0.4561$\pm $0.0972 & 47.7$\pm $1.5 \\
          & PS-xfL1 & 0.0144$\pm $0.0056 & 37.51$\pm $3.47 & 0.9106$\pm $0.0251 & 182.5$\pm $0.7 \\
          & PS-LLR & 0.0378$\pm $0.0135 & 28.97$\pm $3.07 & 0.8138$\pm $0.0489 & 172.8$\pm $1.6 \\
          & PS-SIDWT & 0.0172$\pm $0.0077 & 36.35$\pm $3.92 & 0.9344$\pm $0.0415 & 419.4$\pm $3.5 \\
          & proposed & \textbf{0.0105}$\pm $\textbf{0.0039} & \textbf{40.17}$\pm $\textbf{3.41} & \textbf{0.9563}$\pm $\textbf{0.0183} & \textbf{11.6}$\pm $\textbf{1.9} \\
    \hline
    \end{tabular}
  \label{table_best_lambda}
  }
\vspace{-1em}
\end{table}

The reconstructed images of one data case using the best $\lambda $  are displayed in ~\autoref{retro-4__Lambda__ZOOM__Mmode}. Obviously, when Nkspc is reduced from 6 to 3, the reconstruction error deteriorates significantly. The images reconstructed by PS-LSQR model is totally corrupted by noise. The images reconstructed by PS-LLR model suffer from dark streaking artifacts on the M-mode motion profile, which implies that the image quality fluctuates from frame to frame. The image reconstructed by PS-xfL1 model displays better image quality, but residual dark streaks can still be observed from the M-mode images. The image reconstructed by PS-SIDWT model displays severe smoothing effects on the M-mode images, where the atrium motion profile is blurred. Scattered noise can be perceived on the images reconstructed by PS-xfL1 and PS-SIDWT method. In comparison, the proposed method produces the best visual image quality and the clearest motion profile. Obviously the reconstruction error level of the proposed method is the lowest among all the models, and the error distribution is also more uniform across the image.

\subsection {Prospective Experiments}

\begin{figure*}
\centerline{\includegraphics[width=\columnwidth]{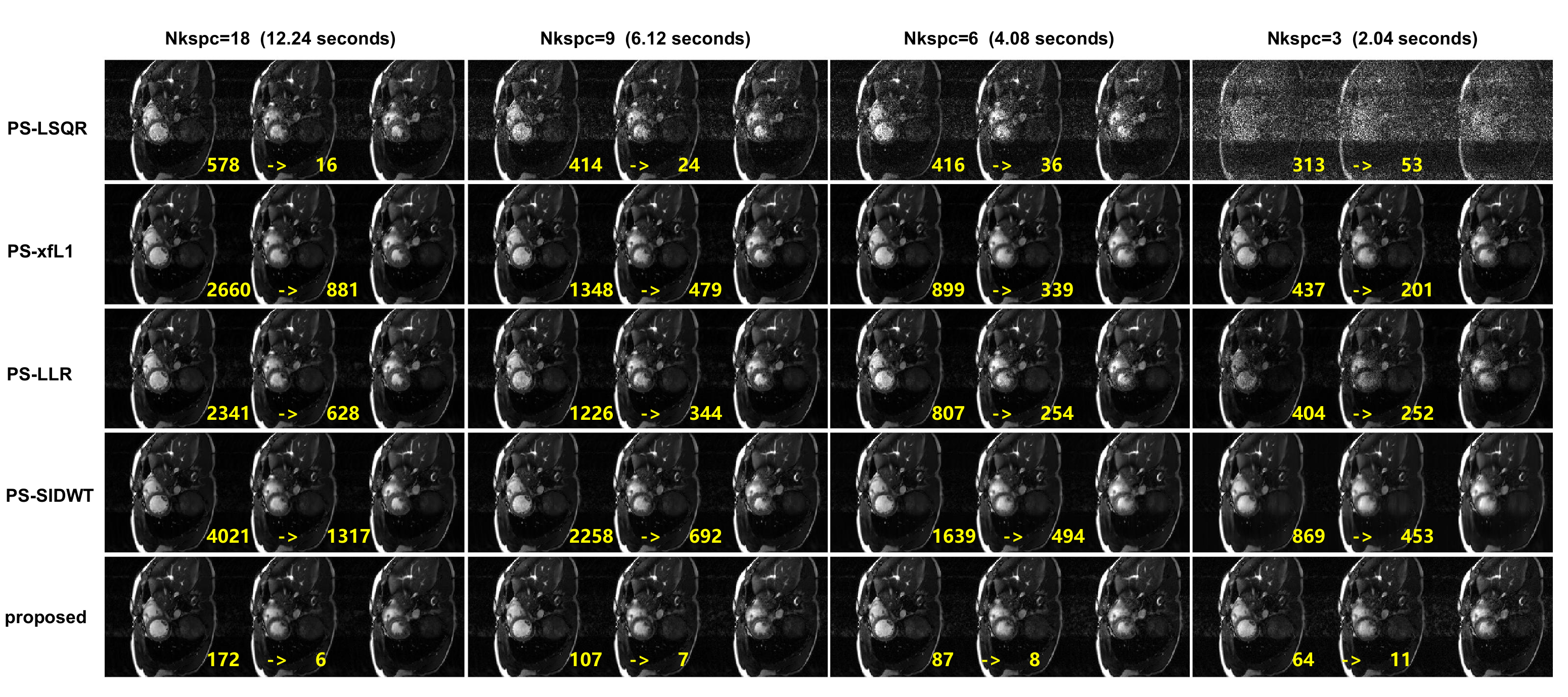}}
\vspace{-0.5em}
\caption{Reconstructed images of a volunteer under Nkspc=18, 9, 6 and 3. For each Nkspc setting, the acquisition time is marked in the bracket. The image frames at the time of diastole, cardiac contraction and systole are displayed. Two reconstruction time is listed at the bottom of each reconstructed image. The reconstruction time without $A^HA$ operator optimization is marked on the left the arrow, while the optimized reconstruction time is marked on the right of the arrow}.
\vspace{-1em}
\label{prosp-1__Nkspc}
\end{figure*}

\begin{figure*}
\centerline{\includegraphics[width=\columnwidth]{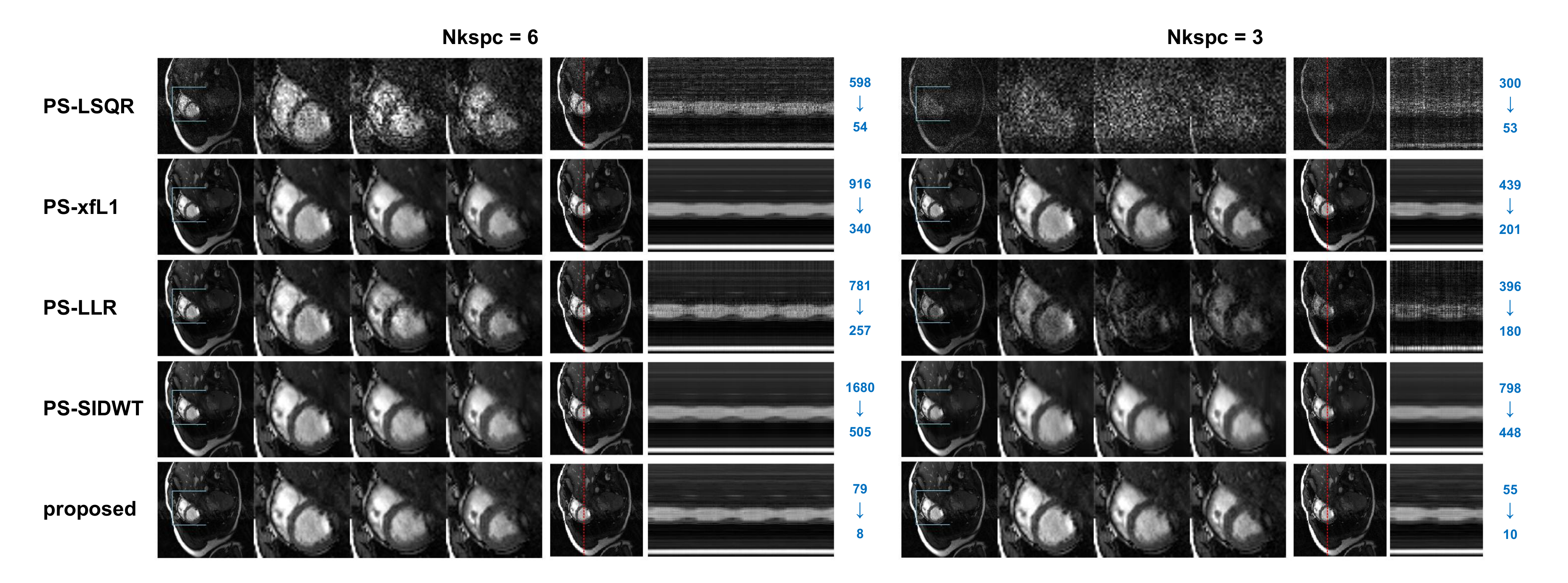}}
\vspace{-0.5em}
\caption{Reconstructed images of one volunteer. The results under Nkspc=6 and Nkspc=3 are compared. For each Nkspc setting, the first column displays the overall reconstructed image. The second, third and fourth column display the zoomed images in the heart region at the time of diastole, cardiac contraction and systole, respectively. M-mode images are plotted along the red dotted line across the heart right atrium.}
\vspace{-1em}
\label{prosp-2__Lambda__ZOOM__Mmode}
\end{figure*}

The reconstructed images of one volunteer are shown in ~\autoref{prosp-1__Nkspc}. The results are consistent with the retrospective experiment, with only slight differences. Under Nkspc=18, all the regularized PS model algorithms have similarly good image quality. However, the image reconstructed by PS-LSQR method is obviously worse than other methods. When Nkspc is further reduced to 9 and 6, the performance of PS-LSQR deteriorates very quickly, while other methods can still maintain relatively good image quality. When Nkspc is reduced to 3, the images reconstructed by PS-LSQR are totally corrupted by noise, PS-LLR method becomes very unstable, the reconstructed images suffer from darkening artifacts. The images reconstructed by PS-SIDWT model are blurred at the blood-myocardium boundary. The images reconstructed by PS-xfL1 model and the proposed method have similar image quality. However, the proposed method is nearly 20-fold faster than the PS-xfL1 algorithm. The zoomed images of another volunteer are shown in ~\autoref{prosp-2__Lambda__ZOOM__Mmode}. The proposed method achieves good image quality and the fastest reconstruction speed simultaneously.

\subsection {Comparative Experiments with non-PS model Methods}

\begin{figure*}
\centerline{\includegraphics[width=\columnwidth]{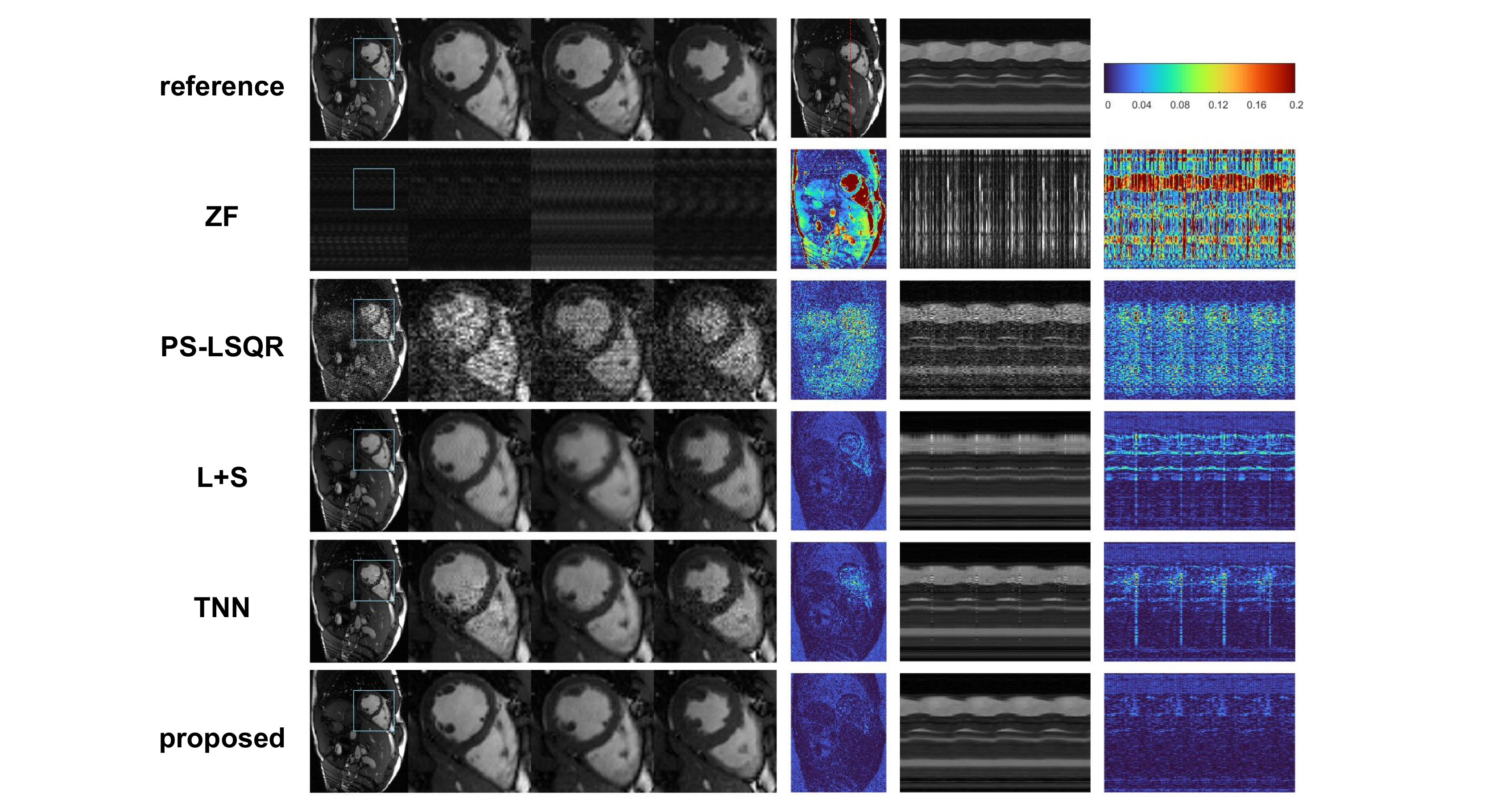}}
\vspace{-0.5em}
\caption{A comparison of the reconstructed image quality with the non-PS model methods. ZF indicates the zero-filling reconstruction method. Because the undersampling rate is very high ($R\approx 42$), the image reconstructed by ZF algorithm has almost no meaningful structures. The image quality reconstructed by the L+S and TNN method is better than the PS-LSQR method. The proposed accelerated PS model method has the best image reconstruction quality.}
\vspace{-1em}
\label{nonPS-1__ZOOM__Mmode_v2}
\end{figure*}

The comparison of the reconstructed images with the non-PS model methods is shown in ~\autoref{nonPS-1__ZOOM__Mmode_v2}. Because the undersampling rate is very high ($R\approx 42$), the images reconstructed by ZF algorithm have almost no meaningful structures. The images reconstructed by PS-LSQR method are severely corrupted by noise. The L+S method has good denoising effect. The images reconstructed by the L+S method have much better visual quality than PS-LSQR method. However, significant over-smoothing effect can be perceived from the images reconstructed by the L+S method. The loss of image details is most obvious at the frame of cardiac contraction, where the blood-myocardium boundary is obscured. The images reconstructed by the TNN method are sharper. Compared with the L+S method, the reconstructed motion profile of the TNN method is more consistent with the reference. However, noise corruption has not been removed clearly for the TNN method. Residual noise-like artefacts can be observed on the zoomed images and M-mode images. The image quality of the proposed accelerated PS model method outperforms other methods significantly. The quantitative metrics of this study are listed in ~\autoref{table_nonPS}, which are consistent with the visual appearance.

\section{Discussion}
\label{sec:discussion}

\subsection{Explanation of Results}

The most important feature of the PS model is the explicit dimension reduction. This property not only can improve the inverse problem conditioning, but can also save reconstruction time. However, most of the previous works neglected this feature and destroyed this property in the algorithm implementation. This is the fundamental cause of long reconstruction time. In this work, we give a detailed analysis of the PS-R model. Two theorems are provided to optimize the $A^HA$ operator, which reduces its computation complexity by $\frac{T}{L}$. Besides, we consciously select the F-norm of temporal difference as the regularization term, and design the optimization objective based on the generalized Tikhonov formulation. Both parts contribute to the acceleration of the PS model. The reconstruction time can be reduced to only a few seconds by the proposed algorithm.

\begin{table}[htbp]
  \centering
  \caption{Quantitative results of the comparative experiments with two non-PS model methods}
  \resizebox{\linewidth}{!}{
    \begin{tabular}{ccccc}
    \hline
     Model & nRMSE & PSNR  & SSIM  & recon time (s) \\
    \hline
    \multirow{5}[2]{*}{}
          ZF & 0.1513$\pm $0.0184 & 16.47$\pm $1.10 & 0.1284$\pm $0.0472 & 0.7$\pm $0.0 \\
          PS-LSQR & 0.0262$\pm $0.0110 & 32.55$\pm $4.29 & 0.8233$\pm $0.0799 & 45.7$\pm $6.9 \\
          L+S & 0.0153$\pm $0.0045 & 37.88$\pm $3.70 & 0.9321$\pm $0.0200 & 840.0$\pm $24.8 \\
          TNN & 0.0138$\pm $0.0029 & 38.89$\pm $2.52 & 0.9370$\pm $0.0184 & 13576.9$\pm $21.5 \\
          proposed & \textbf{0.0080}$\pm $\textbf{0.0030} & \textbf{42.68}$\pm $\textbf{3.72} & \textbf{0.9607}$\pm $\textbf{0.0170} & \textbf{7.9}$\pm $\textbf{1.4} \\
    \hline
    \end{tabular}
  \label{table_nonPS}
  }
\end{table}

This paper is mainly inspired by the previous work\cite{T2-shuffling} which utilized the $A^HA$ operator optimization for accelerating the reconstruction of T2-weighted imaging for knee and feet. Our work has two main differences compared with this work. The first is that the dimension-reduced optimization is not only used in the $A^HA$ operator, but also in the regularization term design. The previous work\cite{T2-shuffling} utilized a time-consuming patch-based nuclear norm regularization term (Locally-Low-Rank), which leads to long reconstruction time. In this work, we fully exploit the dimension-reduced property of PS model, and intentionally design a Tikhonov regularization term, which is shown to be much more efficient in practice. The second difference lies in the dynamic modeling. The previous work is used for anatomical imaging where the object is static while the image contrast is evolving with time. The signal subspace is extracted from Bloch equation simulation. In this work, we extend the optimization technique into a totally different imaging scenario, where the contrast is held stable but multiple physiology motion patterns (heart-beat and free-breathing) are involved. The signal subspace is extracted from additionally acquired navigator echoes. Self-navigator is widely used for extracting the dynamic information adaptively, therefore, the proposed method may be beneficial to many other related imaging applications.

Furthermore, the proposed method also displays superior performance on the image quality in the experiments. The mean value of nRMSE, PSNR and SSIM of the proposed method is systematically higher than the widely-used PS-sparse model. There are two possible reasons for this phenomenon. First, in compressed-sensing theory\cite{Sparse_MRI}\cite{Compressed_sensing_MRI}, successful minimization of data sparsity is based on the incoherence assumption, which requires the sampling pattern to be incoherent with the sparse representation. In this study, Cartesian trajectory is used for the acquisition, which means the under-sampling only occurs on the phase encoding direction. This leads to structured aliasing in the image, which is hard to be removed by enforcing sparsity. The second possible reason may be related to the subspace constraints. Because the subspace is spanned by only a few basis vectors, some sparse solutions of the image may be excluded from the subspace. Therefore, even if the sparsity enforcement is carried out correctly and effectively, the solution may lose the sparsity after being projected onto the subspace, leading to a suboptimal result. The proposed method expresses the PS model in a generalized Tikhonov formulation, which avoids this problem and gives better reconstruction quality.

Because the frequency encoding direction is fully-sampled in Cartesian trajectory, some studies used this property to accelerate the PS model\cite{PSF+xfL1____ADMM____cardiac_imaging____recon_time_11min}. By performing one-dimensional inverse Fourier transform along this direction, the 2D reconstruction problem can be decomposed into several 1D problems. However, this trick is only applicable to Cartesian trajectory. The dimension-reduced technique proposed in this paper can be used in combination with arbitrary trajectories, thus is promising for more advanced imaging techniques.

\subsection{Limitations and Future Work}

One of the most important hyper-parameters for PS model is the model order $L$, which is actually the rank of the temporal subspace. Many previous studies have explored and discussed the impact of model order $L$ on reconstruction quality\cite{PSF_heart_digital_phantom_simulation____select_the_best_model_order____according_to_the_max_difference_of_singular_value}\cite{PSF+xfL1____ADMM____cardiac_imaging____recon_time_11min}. Generally speaking, if $L$ is too small, PS model is unable to express complex dynamic signal. When $L$ is too large, the reconstruction will be sensitive to noise. In this work, the best $L$ is selected by simulation based on a noise statistical model\cite{T2-shuffling}. This method can save the time for doing lots of reconstruction experiments. However, the estimated error is actually a theoretical value, which is based on two underlying assumptions. The first is that the noise is independent Gaussian distribution variable. The second assumption is that the reconstruction algorithm is perfect. Therefore, the reconstruction error in practice will be always higher than this theoretical value. In other words, this method provides a lower bound for the reconstruction error, which is only heuristic for the selection of $L$. Extensive reconstruction experiments are indispensable if a more precise choice of the best $L$ is needed.

Although similar phenomenon can be observed between the retrospective and prospective experiment results, the image quality reconstructed from in-vivo data is worse than the simulation data. There are many reasons accounting for this gap. First, the respiratory motion in the in-vivo data can be more variable than the simulation data. The combination of respiratory motion and heart beat can produce very complicated motion pattern, which is hard to be resolved by linear modeling. This problem may be relieved by extracting non-linear temporal subspace\cite{PSF+kernelPCA_V____dynamic_cardiac_and_perfusion}\cite{PSF_radial____manifold_learning_nonlinear_V____realtime_cardiac} or using high-order PS models\cite{MR-multitasking_cardiac}\cite{PSF_model_extend_to_tensor____breath_and_inversion_recovery____U_and_TV_sparse____ADMM____free_breathing_3D_cardiac_T1mapping}. In this work, we simply use SVD to extract the temporal subspace because it is the most frequently used method. The proposed dimension-reduced optimization technique is actually applicable to any other types of temporal subspace to obtain better results for the in-vivo data. Second, random phase encoding order is used in this paper, because the reconstruction quality is proved to be much better than linear phase encoding order in our preliminary study. However, it has been reported previously that the signal steady state of bSSFP sequence will be disturbed if the phase encoding gradient changes abruptly during acquisition\cite{bSSFP_PE}\cite{bSSFP_3D}. Unstable signal amplitude will cause artifacts in the reconstructed images. In the future, more sophisticated trajectory and phase encoding order may be considered to achieve incoherent sampling and stable signal steady state simultaneously.

\section{CONCLUSION}
\label{sec:conclusion}
In this paper we propose to use dimension-reduced optimization technique to accelerate the partial separable model for dynamic MRI. We have demonstrated that the proposed method outperforms other popular PS models in high temporal resolution free-running cardiac MRI. The proposed method is simple, fast, easy to implement, insensitive to hyper-parameters, and produces better image quality in both simulated and in-vivo data. The proposed method may be helpful for many related time-consuming imaging techniques, and provide new possibilities for more challenging dynamic imaging applications in the future.

\section{Acknowledgements}
The authors would thank Dr. Shuo Chen for his contribution with ideas in the early stage of this work. We also thank Dr. Zhensen Chen for his help in MRI sequence programming. We thank Yichen Zheng for her preliminary studies as the foundation of this work.

\section{Funding}
This work was supported in part by the National Natural Science Foundation of China under Grant 81971604 and in part by the Natural Science Foundation of Beijing Municipality under Grant L192013. 

\section{Appendix}

\subsection{Theorem 1 operator exchange-ability}

In this section, we will prove that the computation order of operator $V$ can be exchanged with $S$ and $F$, respectively. In other words, we will prove that the following two implementations are equivalent:

\begin{equation}
A(U)=MFSVU=MVFSU.
\label{A_equation}
\end{equation}
It should be noted that the expression in \eqref{A_equation} is a simplified form, which only considers the operator order. The real computation implementations for $A$ operator is written as follows:

\begin{equation}
A(U)=MFSVU=M\odot (F\cdot (S\odot (U\cdot V))),
\label{A_equation_real}
\end{equation}
where $\cdot $ denotes the matrix multiplication, and $\odot $ denotes the element-wise multiplication. Therefore, this theorem is actually written as:

\begin{equation}
M\odot (F\cdot (S\odot (U\cdot V))) = M\odot ((F\cdot (S\odot U))\cdot V).
\label{T1_real}
\end{equation}
The equation \eqref{T1_real} is equivalent to the following two equations:

\begin{equation}
S\odot (U\cdot V) = (S\odot U)\cdot V,
\label{S_exchange}
\end{equation}
\begin{equation}
F\cdot (SU\cdot V) = (F\cdot SU)\cdot V.
\label{F_exchange}
\end{equation}

Because of the associative law of matrix multiplication, the exchange-ability of equation\eqref{F_exchange} is obviously correct. Besides, for each coil channel, the sensitivity map operator can be formulated into a diagonal matrix $S_j\in C^{N\times N}$, where the diagonal elements are the spatial sensitivities.
\begin{equation}
S_j\cdot (U\cdot V) = (S_j\cdot U)\cdot V.
\label{Sj_exchange}
\end{equation}
Therefore, the exchange-ability between $S_j$ and $V$ is correct. Because $S$ is just stacked by $S_j (j=1,\cdots ,J)$, the exchange-ability between $S$ and $V$ is also correct.

In conclusion, the computation order of $V$ can be exchanged with $S$ and $F$ respectively. Theorem 1 is proved. In fact, this theorem can be generalized to any spatial encoding operator which is independent of the time dimension, such as Non-uniform FFT. Therefore, this technique is applicable to any other non-uniform k-space sampling trajectory.

\subsection{Theorem 2 operator merge-ability}

\begin{figure}
\centerline{\includegraphics[width=0.5\columnwidth]{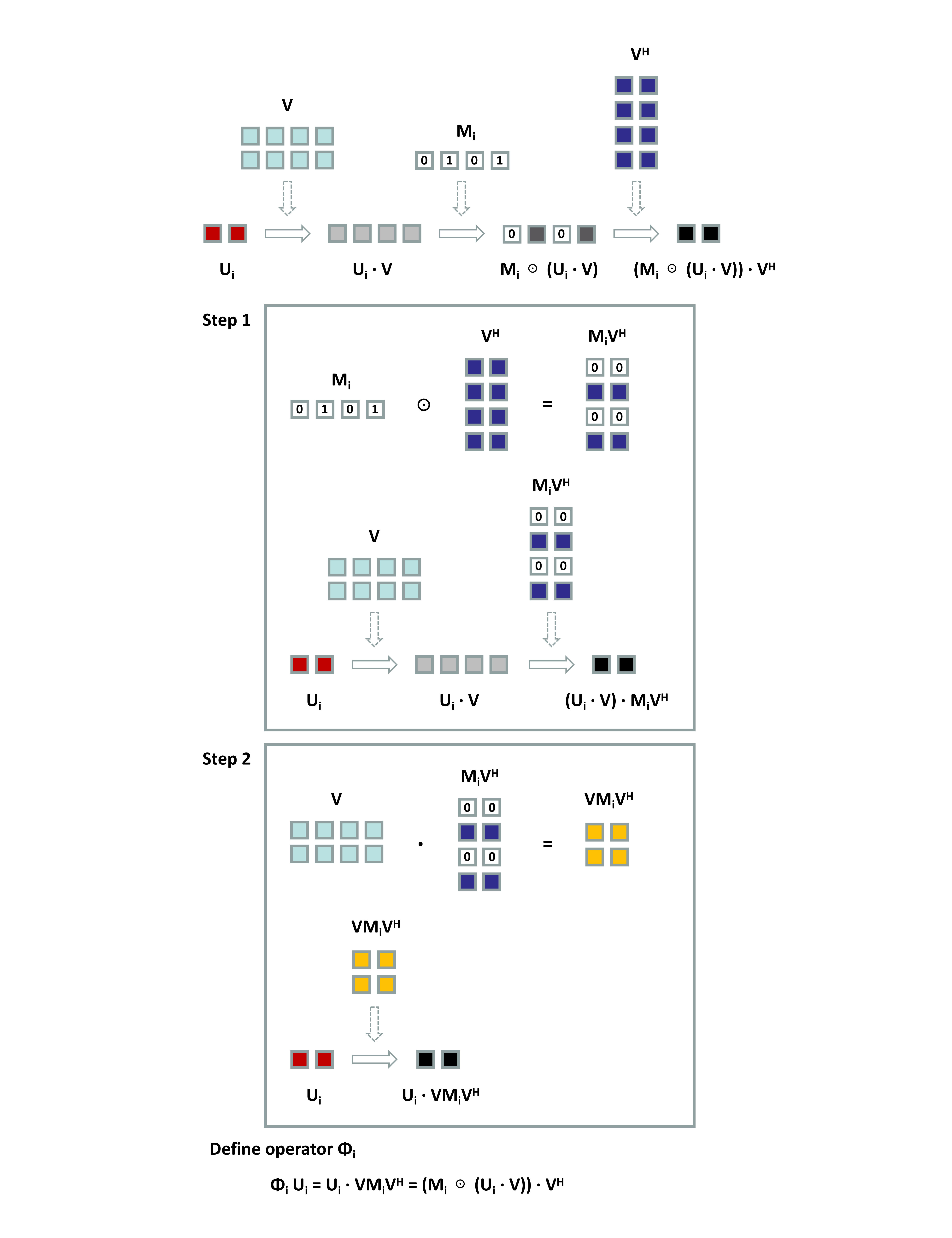}}
\vspace{-0.5em}
\caption{Illustration of the Theorem 2 about operator merge-ability. The computation can be decomposed into 1D problems row by row, so we need only to prove the case for the i-th row (i=1,...,N). $U_i$ denotes the i-th row of $U$, and $M_i$ denotes the i-th row of $M$, $V$ denotes the temporal basis functions. For simplifying the illustration, the diagram displays the proof for $L=2$ and $T=4$.}
\vspace{-1em}
\label{VHMV_PDF}
\end{figure}

In this section, we will prove that the three operators $V^H$, $M$ and $V$ can be merged into a single equivalent operator $\Phi $, which only operates in the L-dimensional space:

\begin{equation}
V^HMVFSU=\Phi FSU.
\label{merge}
\end{equation}
Without loss of correctness, equation\eqref{merge} can be simplified to:
\begin{equation}
V^HMVU=\Phi U.
\label{merge_U}
\end{equation}

Similar to the proof of Theorem 1, the expression in equation\eqref{merge_U} is a simplified form, which only considers the operator order. Rewrite equation\eqref{merge_U} to its real computation form:
\begin{equation}
V^HMVU = (M\odot (U\cdot V))\cdot V^H,
\label{merge_U_real}
\end{equation}
where $U\in C^{N\times L}$, $V\in C^{L\times T}$, $M\in R^{N\times T}$.  Obviously, equation\eqref{merge_U_real} can be separated along the spatial dimension. Therefore, we only need to consider the computation in the i-th row:
\begin{equation}
V^HM_iVU_i=(M_i\odot (U_i\cdot V))\cdot V^H,
\label{V^HMVU_i}
\end{equation}
where $U_i$ denotes the i-th row of $U$, and $M_i$ denotes the i-th row of $M$. The diagram in ~\autoref{VHMV_PDF} provides an schematic proof for equation\eqref{V^HMVU_i}. It can be seen that $V^HM_iV$ can be merged into a single operator $\Phi _i \in C^{L\times L}$, which satisfies the computation equivalence:
\begin{equation}
\Phi _iU_i=V^HM_iVU_i=(M_i\odot (U_i\cdot V))\cdot V^H.
\label{Phi_i}
\end{equation}

It should be noted that for each row location $i$, the undersampling mask $M_i$ is different. Therefore, different $\Phi _i$ will be generated for each row, constituting an operator $\Phi \in C^{N\times L\times L}$, which is exactly the equation for Theorem 2:
\begin{equation}
\Phi U=V^HMVU=(M\odot (U\cdot V))\cdot V^H.
\label{merge_final}
\end{equation}

\bibliographystyle{elsarticle-num} 
\bibliography{reference}





\end{document}